\documentclass[twocolumn]{aastex631}

\usepackage{comment}

\usepackage{{amsmath}}

\newcommand{\vect}[1]{{\mathbf{{#1}}}}
\newcommand{\uv}[1]{{\hat{\mathbf{e}}_{#1}}}
\newcommand{\pder}[2]{\frac{\partial#1}{\partial#2}}
\newcommand{\oder}[2]{\frac{\mathrm{d}#1}{\mathrm{d}#2}}

\newcommand{\rsun}{\mathrm{r}_{\odot}}

\newcommand{\krad}{\kappa}

\shorttitle{Turbulence in Parker Spiral geometry}
\shortauthors{Laitinen et al.}
\graphicspath{{./}{figures/}}

\begin{document}

\title{An analytical model of turbulence in Parker spiral geometry and associated magnetic field line lengths}



\author[0000-0002-7719-7783]{T. Laitinen}
\affiliation{Jeremiah Horrocks Institute \\
University of Central Lancashire\\
UK}

\author[0000-0002-7837-5780]{S. Dalla}
\affiliation{Jeremiah Horrocks Institute \\
University of Central Lancashire\\
UK}

\author[0000-0003-4390-2920]{C. Waterfall}
\affiliation{Jeremiah Horrocks Institute \\
University of Central Lancashire\\
UK}

\author[0000-0002-9362-7165]{A. Hutchinson}
\affiliation{Jeremiah Horrocks Institute \\
University of Central Lancashire\\
UK}

\begin{abstract}
Understanding the magnetic connections from the Sun to interplanetary space is crucial for linking in situ particle observations with the solar source regions of the particles. A simple connection along the large-scale Parker spiral magnetic field is made complex by the turbulent random-walk of field lines. In this paper, we present the first analytical model of heliospheric magnetic fields where the dominant 2D component of the turbulence is transverse to the Parker spiral. The 2D wave field is supplemented with a minor wave field component that has asymptotically slab geometry at small and large heliocentric distances. We show that turbulence spreads field lines from a small source region at the Sun to a 60$^\circ$ heliolongitudinal and -latitudinal range at 1~au, with standard deviation of the angular spread of the field lines $14^\circ$. Small source regions map to an intermittent range of longitudes and latitudes at 1~au, consistent with dropouts in solar energetic particle intensities. The lengths of the field lines are significantly extended from the nominal Parker spiral length of 1.17~au up to 1.6~au, with field lines from sources at and behind the west limb considerably longer than those closer to the solar disk centre. We discuss the implications of our findings on understanding charged particle propagation, and the importance of understanding the turbulence properties close to the Sun.
\end{abstract}

\keywords{Interplanetary turbulence (830), Interplanetary physics (827), Heliosphere (711)}

\section{Introduction} \label{sec:intro}

Interplanetary space consists of plasma that flows from the Sun as solar wind and extends to the outer heliosphere. This plasma is permeated by a magnetic field which extends as open field from the Sun and, due to solar rotation, forms an Archimedean spiral, the so-called Parker spiral \citep{Parker1958}. Fast charged solar particles are guided along the magnetic field, and through in situ observations of these particles we can gain information on processes that heat and accelerate particles at the Sun via different mechanisms from thermal to relativistic energies.

The analysis of in situ particle observations and comparison to the solar sources, however, is complicated by the presence of fluctuations of order up to the background magnetic field magnitude in the solar wind \citep[e.g.][and references therein]{Bruno2005}. Analysis of heliospheric magnetic field observations \citep[e.g.][]{Matthaeus1990, Bieber1996} suggests that the majority of the fluctuations are transverse, varying in the direction normal to the background magnetic field. Such turbulence will cause the field lines to meander across the mean field direction, either circling helically around magnetic islands, or spreading through x-points between such islands \citep[e.g][]{Chuychai2007}. This "2D landscape" is broken by variation of the fields in the direction along the background field caused by the nonlinear evolution of the turbulence \citep{GoSr1995}.

Turbulence in magnetised plasmas has been investigated using various computational approaches \citep[see, e.g.][for review]{Beresnyak2019_MHDturbreview}. Plasma turbulence simulations are, however, computationally very expensive, and for applications where the magnetic connectivity and its effects on plasma and particle transport are investigated, often simpler methods are used. A popular choice is to describe plasma turbulence as a superposition of random-phase transverse wave modes with wavenumber vectors normal (the so-called 2D component) and along (the slab component) the background magnetic field. This so-called composite turbulence model has been widely used for the analysis of field lines \citep[e.g.][]{Matthaeus1995, Ruffolo2004, Chuychai2007} as well as cosmic rays \citep[e.g.][]{GiaJok1999, Qin2002, Tautz2010, LaEa2012}, often comparing numerical simulations with theoretical descriptions. In most of these models, spatially homogeneous turbulence is superposed on a uniform and constant background magnetic field. Such a configuration is useful for determining how the field line or particle diffusion coefficients depend on the properties of the turbulence, and the temporal scales of the charged particle propagation along the meandering field lines.

However, the homogeneous environment and assumption of uniform constant background magnetic field $\vect{B}_0$ fails to properly account for the interplay between the spatially varying turbulence \citep[e.g.][]{Bruno2005} and the large-scale Parker spiral magnetic field. We are often interested in the propagation of charged particles from the Sun to the Earth, or other locations in the heliosphere where observations are made with in situ instruments. Stochastic processes are likely to alter the field line connection between the solar source and interplanetary point of observation, and to extend the length of the field lines the particles propagate on. However, the connectivity and the lengthening of the field lines will depend also on the underlying Parker spiral geometry that introduces asymmetry to the field line configuration. Will such asymmetry result also in asymmetry of the path lengths from the sun to 1~au, and to what extent are the field line lengths extended? What properties of the heliospheric turbulence affect the path length, and how?

Aside of diffusion-based and random walk models \cite[e.g.][]{PommoisEa2001, PommoisEa2001_JGR,LaDa2019pathlength,Chhiber2021_randomwalk, BianLi2022}, three main approaches have been used to investigate turbulent magnetic field lines in a heliospheric configuration. \citet{Giacalone2001} introduced a model where the motion of the magnetic field footpoints in the photosphere is modelled with a stream function at the solar surface, resulting in magnetic fluctuations in the latitude-longitude plane. The \citet{Giacalone2001} model benefits from the physical link between the photospheric motions and the interplanetary turbulence, thus providing an observational constraint for the source of the turbulence. However, the fluctuations in the model are transverse at the solar surface, whereas to remain transverse in the Parker spiral geometry, the fluctuations would also need a radial component of the fluctuating field. Thus the \citet{Giacalone2001} model does not fulfil the requirement of the transverse nature of the plasma turbulence modes in the outer heliosphere. \citet{RuffoloEa2013} used a different approach, where the composite turbulence model with constant background magnetic field is projected on a spherically extending sector. This enabled their model to be fully transverse, with 2D and slab modes with respect to the radial magnetic field. However, their model is limited to a radial background magnetic field and thus lacks the asymmetry that is created by the Parker spiral. Finally, \citet{Tautz2011} and \citet{Fraschetti2018} superposed isotropic turbulence on Parker spiral geometry, however, the isotropic turbulence model does not agree with the expected dominance of transverse 2D modes in the heliosphere, and, as noted by \citet{Fraschetti2018}, the turbulent magnetic field in their model was not divergence-free. The difficulty of keeping the magnetic field divergence-free is the crux of the problem: in Parker spiral geometry, typically presented in heliocentric spherical coordinates $(r, \theta, \phi)$, the magnetic field vectors of transverse fluctuations, $\delta\vect{B}$, in general have a non-vanishing radial component, which in the general case depends on the radial coordinate. Such component, $\delta B_r(r)$ generates divergence which must be cancelled by $\theta$- or $\phi$-dependence of the $\theta$- or $\phi$-component of the fluctuating vector. 

In this paper, we introduce a novel analytical model for interplanetary turbulence, where the background magnetic field is of Parker spiral shape, and $\nabla\cdot\vect{B}=0$ everywhere. The turbulence is dominated by 2D mode waves for which the fluctuating vector is normal to the Parker spiral, and the wave vector is normal to both the Parker spiral and the fluctuating vector $\delta\vect{B}$. The 2D mode turbulence is supplemented with a minor component of slab mode waves, which are approximated as a superposition of radial and azimuthal waves. The model allows us to calculate analytically the turbulent magnetic field everywhere in the heliosphere, and the logarithmic spacing of the wave modes allows us to cover a large range of turbulence scales with small number of wave modes. Using the newly-constructed model, we investigate the meandering of the field lines across the Parker spiral, comparing our results with recent results that evaluate field-line random walk as diffusion. We apply our model to evaluate the length of field lines from the Sun to 1 au, demonstrating how the asymmetry created by the Parker spiral is reflected as asymmetry in the field line lengths. We present the turbulence model in Section \ref{sec:models}, and our results in Section~\ref{sec:results}. We discuss the significance of our results in Section~\ref{sec:discussion} and draw our conclusions in Section~\ref{sec:conclusions}.

\section{Turbulence model} \label{sec:models}

\subsection{2D turbulence} \label{sec:2D}

We model the 2D magnetic field turbulence in Parker spiral geometry as a superposition of Fourier modes with both the fluctuating vector
$\delta\vect{B}$ and the wave number vector $\vect{k}$ normal to the background magnetic field direction and $\delta\vect{B}\perp \vect{k}$. In addition, the divergence-free condition $\nabla\cdot\delta\vect{B=0}$ must be fulfilled. These requirements can be satisfied by introducting a vector potential $\vect{A}$ which is aligned with the background magnetic field and has a phase term that is constant along the background field. The transverse fluctuating field can then be obtained from this vector potential as $\delta\vect{B}=\nabla\times\vect{A}$. In Cartesian geometry with coordinates $(x, y, z)$ and constant background magnetic field directed along the $z$-coordinate unit vector $\uv{z}$, a transverse 2D mode with magnetic field amplitude $\delta B_0$ can be constructed with a vector potential $\vect{A}=\delta B_0 k^{-1}\sin\left(g(x,y)\right)\uv{z}$, where $g(x,y)$ is the oscillating phase of the wave, given by $\vect{k}\cdot\vect{r}+\delta$ where $\delta$ is a random phase offset and $\vect{r}$ a cartesian position vector. 
Choosing $\mathbf{k}=-k\cos{\alpha}\,\uv{x}+k\sin\alpha\,\uv{y}$, with  polarisation angle $\alpha$, we get fluctuating field components 

\begin{equation*}
    \delta\vect{B}_i=\delta B_{0i} \sin\alpha\sin\left[g(x,y)\right]\uv{x}+\delta B_{0i} \cos\alpha\sin\left[g(x,y)\right]\uv{y}.
\end{equation*}

To construct the vector potential in spherical geometry, \citet{RuffoloEa2013} suggested a conversion of the Cartesian Fourier modes to heliolongitude $\phi=x/r$ and heliolatitude $\Lambda=y/r$. We use the same approach, and cast the phase of the 2D waves in the form 

\begin{equation*}
    g(\theta, \phi) = \krad\phi \sin\alpha-\krad \theta\cos\alpha+\delta
\end{equation*}
where $\theta=\pi/2-\Lambda$. Note that in this representation, the wave vector $\krad=k r$ is now in angular units $\mathrm{rad}^{-1}$.

We define a vector potential 
\begin{equation}
\vect{A}_i=\delta B_{0i}(r, \krad) \krad^{-1}\sin\left(g(\theta,\phi)\right)\uv{r}    
\end{equation}
from which we obtain the fluctuating magnetic field vector
\begin{multline*}
    \delta\vect{B}_i=\frac{1}{r\sin\theta}\delta B_{0i}(r,\krad) \sin\alpha\sin\left[g(\theta,\phi)\right]\uv{\theta}+\\ \frac{1}{r}\delta B_{0i}(r,\krad) \cos\alpha\sin\left[g(\theta,\phi)\right]\uv{\phi}.    
\end{multline*}

In Parker spiral geometry, we must take into account that the 2D modes must have their wave vectors normal to the Parker spiral direction, 

\begin{equation}
    \hat{\mathbf{B}}=\cos\psi \uv{r} - \sin\psi \uv{\phi}
\end{equation}
where $\psi$ is the angle between the radial direction and the Parker spiral, and 

\begin{equation*}
    \cos\psi=\frac{1}{\sqrt{1+(r \sin\theta/a)^2}}
\end{equation*}
where $a=v_{sw}/\Omega_\odot$. For a 2D mode in this geometry, we introduce a phase term

\begin{equation*}
    g(r, \theta, \phi) = (k_r r+\krad \phi)\sin\alpha+\krad \theta\cos\alpha
\end{equation*}
where the radial wavenumber $k_r$ is chosen so that for a displacement from $(r, \theta, \phi)$ to $(r+\delta r, \theta, \phi+\delta\phi)$  along the Parker spiral the phase term remains constant, that is

\begin{equation*}
    (k_r \delta r+\krad \delta\phi)\sin\alpha=0.
\end{equation*}
A displacement $\delta s$ along the Parker spiral has components $\delta r = \delta s \cos\psi$, $r\sin\theta \delta\phi = -\delta s \sin\psi$. Thus, for constant phase over the displacement we find

\begin{equation*}
k_r r=\frac{ \tan\psi}{\sin\theta}=\krad \frac{r}{a},
\end{equation*}
and the phase term can be written as
\begin{equation}
    g(r, \theta, \phi)=\left(\krad\frac{r}{a}+\krad \phi\right)\sin\alpha+\krad \theta\cos\alpha+\delta.
\end{equation}
Subsequently, the vector potential can be written as 
\begin{multline} \label{eq:2dvecpot}
     \vect{A}_i=\delta B_{0i}(r, \krad) \krad^{-1} \\
     \sin\left[\left(\krad\frac{r}{a}+\krad \phi\right)\sin\alpha+\krad \theta\cos\alpha+\delta\right] \\
     \left(\cos\psi \uv{r} - \sin\psi \uv{\phi}\right).
\end{multline}
The magnetic field can be obtained from this vector potential trivially as $\delta\vect{B}_i=\nabla\times\vect{A}_i$.  However, with the introduction of radial dependence of the phase term and the azimuthal vector potential component, the expression for $\delta\vect{B}_i$ is lengthy and will not be written down here.

Following \citet{GiaJok1999}, we define a spectrum of wave modes of logarithmically spaced wave numbers between $\krad_0$ and $\krad_1$ as
\begin{equation}\label{eq:2dspectrum}
    \delta B_{0i}^2 = \delta B_0^2(r) \,C\frac{2\pi \krad_i \Delta \krad_i\;\krad_i^p}{1+(\lambda_{c\perp}(r)\, \krad_i/r)^{q+p}}
\end{equation}
where $\krad_i$ is the i'th wave number, $\Delta \krad_i$ the corresponding width of the wavenumber bin and $C^{-1}=\sum_i \delta B_{0i}^2$ a normalisation factor. The spectral shape is composed of an inertial range with Kolmogorov spectral shape given by $q=8/3$ at large wave numbers, and the so-called energy-containing range with spectral index $p$ at low wave numbers. The transition between the two scales takes place at break scale length $\lambda_{c\perp}(r)$. The variance of the spectrum is given by $\delta B_0^2(r)$. The division of $\krad_i$ by $r$ in the denominator is done to take into account the fact that $\krad_i$ is in units rad$^{-1}$, whereas the spectral break scale $\lambda_{c\perp}(r)$ is defined in length units. Note that both $\lambda_{c\perp}(r)$ and $\delta B_0(r)^2$ have radial dependence, which must be taken into account when taking $\nabla\times\vect{A}$.

We introduce a cyclic boundary condition in $\phi$, which gives a condition for the wavenumber $\krad=n/\sin\alpha$ for $\alpha\neq 0$ and integer $n$. We do not introduce boundaries in the $\theta$ direction, thus our model is not applicable across the poles.

\begin{figure}
\plotone{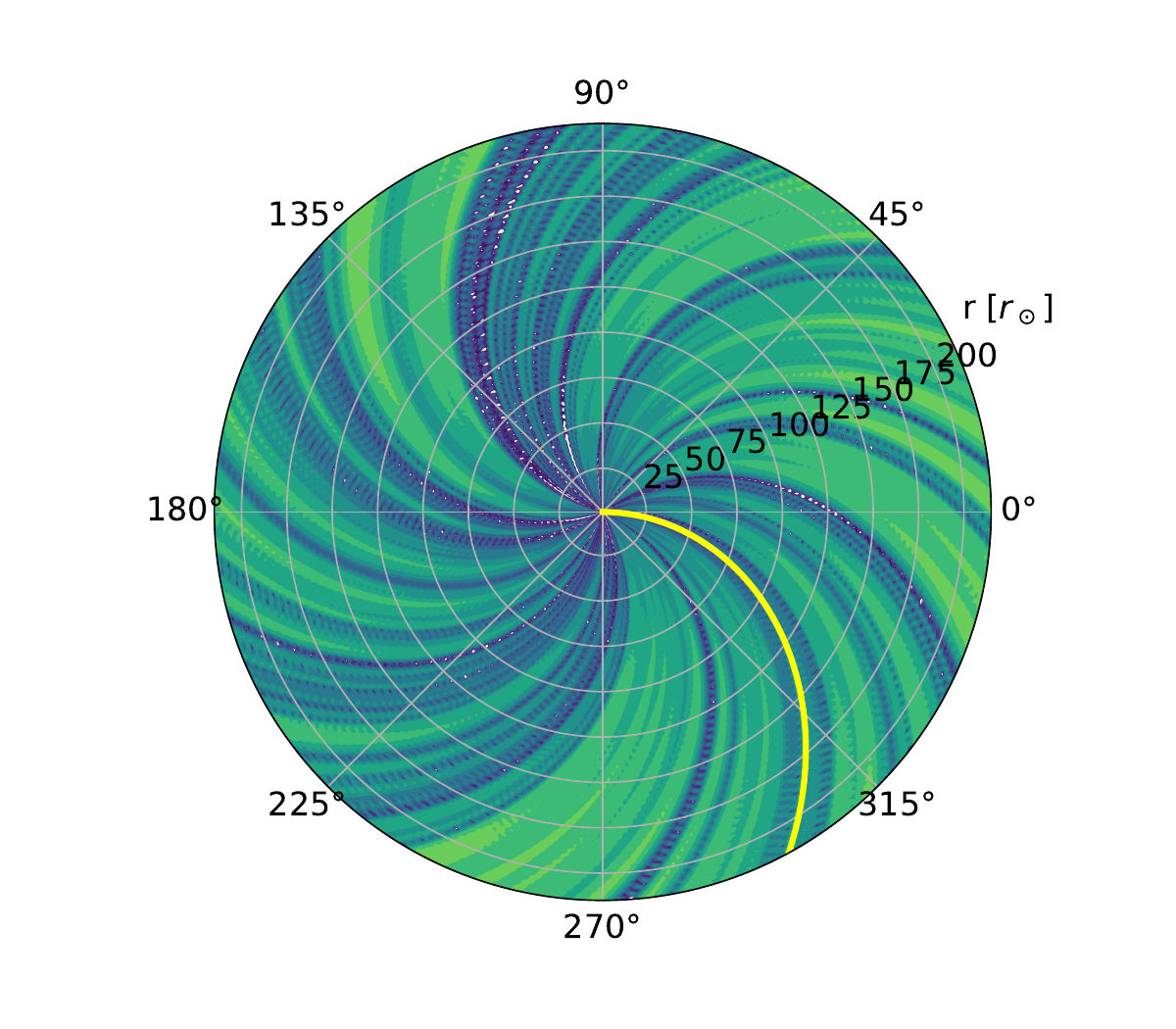} 
\caption{Vector potential magnitude in heliospheric equatorial plane ($\theta=\pi/2$)for 2D turbulence model. The contoured quantity is $\mathrm{log}(|A/\delta B_0(r)|)$. The yellow curve traces the Parker spiral from heliolongitude $\phi=0^\circ$.\label{fig:vecpot_rphi}}
\end{figure}
  
To demonstrate that the turbulence model generated by Equation~(\ref{eq:2dvecpot}) fulfils the requirements of 2D turbulence in Parker spiral geometry, we show the vector potential, Equation~(\ref{eq:2dvecpot}), in Figure~\ref{fig:vecpot_rphi} as a contour of $\mathrm{log}(|A(r, \theta,\phi)/\delta B_0(r)|)$ in the $r-\phi$-plane at $\theta=\pi/2$. As can be seen, the contours trace the Parker spiral, shown by the yellow curve.

\begin{figure}
\plotone{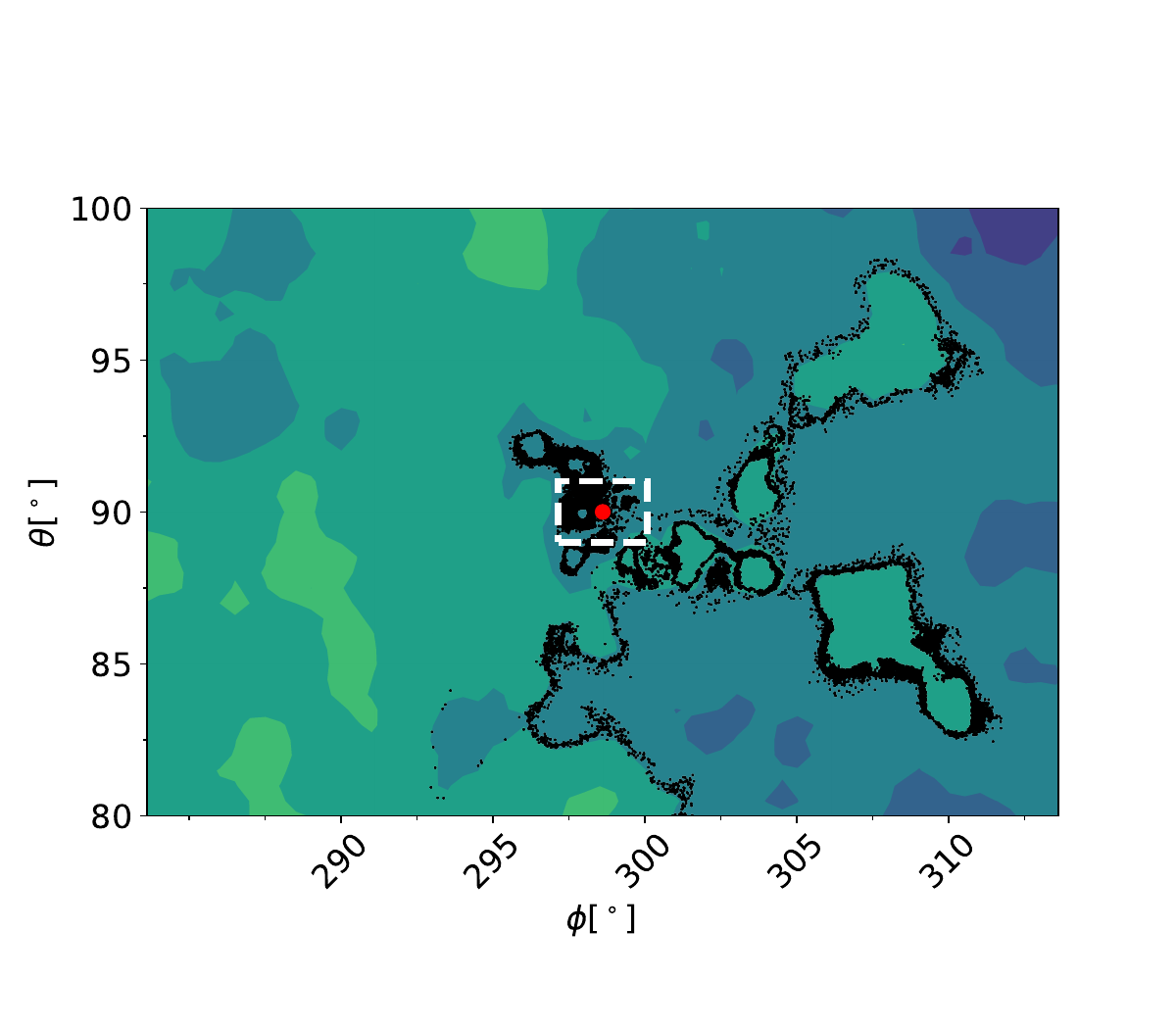} 
\caption{Vector potential at 1 au (coloured contours) for a realisation of the 2D turbulence model. The black dots show where field lines integrated from a $2^\circ\times2^\circ$ area in longitude and latitude at $(r, \theta,\phi)= (2 r_\odot, 90^\circ, 0^\circ)$ cross the 1~au sphere. The white dashed rectangle shows the field line source region translated along Parker spiral. The point where the Parker spiral from the source region centre crosses the 1 au sphere is shown with a red filled circle.\label{fig:vecpot_phitheta}}
\end{figure}

As an additional test of our model, we also inspect whether the magnetic field lines remain on the equipotential surfaces of the vector potential, as expected for 2D turbulence. To this end, we have integrated a sample of 100,000 field lines from an area of $2^\circ\times2^\circ$ in longitude and colatitude centred at $(r, \theta,\phi)= (2 r_\odot, 90^\circ, 0^\circ)$, where $\rsun$ is the solar radius, to 1~au by solving the field line equations 
\begin{equation}\label{eq:fieldline}
\oder{r_i}{s}= \frac{B_i}{B}
\end{equation}
where $r_i$ indicates the cartesian coordinates $x,\,y,\,z$, $B$ is the total magnitude of magnetic field , including the Parker spiral background field, and $s$ is a parameter of the curve in parametric form. In Figure~\ref{fig:vecpot_phitheta}, the black dots indicate where the field lines in the sample cross a 1-au heliocentric sphere, with the white dashed rectangle depicting the mapping of the source region onto 1 au sphere when a simple Parker spiral is used. The dots are superposed on a contour of the vector potential magnitude at 1 au. As can be seen, the field lines cross the 1-au sphere along narrow lanes, which coincide with the vector potential equipotentials, thus indicating that as the field lines traverse the interplanetary space, they remain on the same equipotential surface.

\subsection{Slab turbulence} \label{sec:slab}

A slab mode wave in magnetised plasma is defined as having its wave number vector aligned with the background magnetic field, while the fluctuating vector is in the plane normal to the background field. It is possible to construct such a slab component in Parker spiral configuration using a similar approach as for the 2D component (Section~\ref{sec:2D}). However, the periodic boundary condition in $\phi$ limits $\krad$ to integer multiples of $a/r_\odot\sim 200$, which means that the largest scales in the slab spectrum would be of order $r_\odot/200$, as opposed to largest scales being of order $r_\odot$ for the 2D turbulence. To avoid such disparity of the largest scales in the slab and 2D components, we do not use pure slab turbulence in our Parker spiral turbulence model, but instead use a combination of radial and azimuthal slab modes, with the former dominating in the inner heliosphere and the latter in the outer. The transition between the components is facilitated by multiplying the radial slab component with $\cos\psi'$ and the azimuthal slab component with $\sin\psi'$, where $\psi'\equiv\psi(\theta=\pi/2)$\footnote{The dependence on $\theta$ is removed to avoid unwanted wave modes when taking $\vect{B}=\nabla\times\vect{A}$}.

The radial slab component can be formulated with a vector potential with direction $\hat{\vect{A}}=\uv{\theta}\cos\alpha+\uv{\phi}\sin\alpha$, and with a phase term $g(r)=k_i r+\delta$ that depends only on the radial coordinate $r$. However, because the spectral break point, as well as the scales resonant with energetic particles, evolve with radial distance from the Sun, constant wavenumbers for the discrete wave modes, $k_i$, would remain significant for the simulations only for a limited range in the heliosphere. For this reason, we define the wavenumber magnitude as radially dependent\footnote{The square root dependence is a compromise: using linear dependence would result in constant phase for the radial slab component.}, as $k=k(r)=\krad_i/\sqrt{r a}$. . The phase term is then $g(r)=\krad_i\sqrt{r/a}+\delta$. With this definition, for a given $\krad$ in angular units, $k(r=1\;r_\odot)=\krad/\sqrt{215}\; \mathrm{au}^{-1}$ and $k(r=1\, \mathrm{au})=\krad\; \mathrm{au}^{-1}$.

As the radial slab vector potential described above depends only on the radial coordinate, the radial component of $\nabla\times\vect{A}$ vanishes. We can further simplify the derivation by noting that for a magnetic field vector $B$ that has no radial component $\nabla\cdot\left(C(r)\vect{B}\right)=0$ for an arbitrary $C(r)$. This enables us to first calculate the phase-dependence of the radial slab component field by defining a reduced vector potential 
\begin{align}\label{eq:A_r_rad}
A'_{r, rad} &= 0 \\
A'_{\theta, rad}(r) &=\frac{a'}{r\,\krad\sin\theta}\sin\left(\krad\sqrt{\frac{r}{a'}}+\delta\right)\cos\alpha\\
A'_{\phi, rad}(r) &=\frac{a'}{r\,\krad\sin\theta}\sin\left(\krad\sqrt{\frac{r}{a'}}+\delta\right)\sin\alpha.
\end{align}
and then scaling its curl $\nabla\times\vect{A}'$ appropriately to obtain fluctuating magnetic field with amplitude $\delta B_{0i,rad}(r, \krad)$ as 
\begin{equation}
    \delta \vect{B}_{i, slab, rad}=2 \delta B_{0i,rad}(r, \krad) \cos(\psi')\, r \sqrt{r a'}\; \nabla\times\vect{A'_{rad}}.
\end{equation}
Note that we have also divided the vector potential by $r\sin\theta$ to avoid a radial component arising from the radial component of $\nabla\times\vect{A}'$.

The azimuthal slab turbulence is formed from a vector potential
\begin{align}
A_{r, az}(r, \phi, \krad) =& r\,\sin\theta\, \delta B_{0i,az}(r, \krad)\krad^{-1} \\
& \qquad\qquad\sin\left(\krad\,\phi+\delta\right)\sin\psi'\sin\alpha \nonumber \\ 
    A_{\theta, az}(r, \phi, \krad) =&r\,\sin\theta\, \delta B_{0i,az}(r, \krad)\krad^{-1} \\
    & \qquad\qquad\sin\left(\krad\,\phi+\delta\right)\sin\psi'\cos\alpha  \nonumber \\
    A_{\phi, az} =& 0.\label{eq:A_phi_az} 
\end{align}
It should be noted that because of the $r$ and $\theta$-dependencies of this vector potential, the azimuthal slab mode fields have a non-vanishing $\phi$ component.

The slab spectrum for the slab  modes is given as 
\begin{align}
    \delta B_{i, rad}^2 &= \delta B_0^2(r) \,C_{rad}\frac{\Delta \krad_i\; \krad^p}{1+(\lambda_{c\parallel}(r)\, \krad_i/\sqrt{a r})^{q+p}}  \label{eq:radslabspectrum}\\
    \delta B_{i, az}^2 &= \delta B_0^2(r) \,C_{az}\frac{\Delta \krad_i\; \krad^p}{1+(\lambda_{c\parallel}(r)\, \krad_i/r)^{q+p}} \label{eq:azslabspectrum}
\end{align}
where $\krad_i$ is the i'th wave number, $\Delta \krad_i$ the corresponding width of the wavenumber bin and $C^{-1}=\sum_i dB_i^2$ a normalisation factor. Similar to the 2D spectrum, the spectral shape of the slab spectrum consists of a spectral breakpoint between the flatter low-wavenumber spectrum and the Kolmogorov spectrum, with $q=5/3$, separated by the spectral break scale $\lambda_{c\parallel}(r)$, and the variance of the turbulence is given by $\delta B_0(r)^2$. As in the case of the 2D spectrum, the division of $\krad_i$ by $\sqrt{a r}$ and $r$ in the denominator of Equations~(\ref{eq:radslabspectrum}) and~(\ref{eq:azslabspectrum}), respectively, is done to take into account the fact that $\krad_i$ is in units rad$^{-1}$.

\subsection{Turbulence parameters}\label{sec:turbparams}

The new model of turbulence given by Equation~(\ref{eq:2dvecpot}) for the 2D component and Equations~(\ref{eq:A_r_rad})--(\ref{eq:A_phi_az}) for the slab component is used to simulate the heliospheric turbulent environment in Parker spiral geometry. We construct realisations of our turbulence model as a superposition of typically 1024 logarithmically-spaced 2D and slab mode waves with random polarisations and phases, ranging from the largest scales, $\krad_0=1$ to $\krad_1=100,000$. For the 2D spectrum and the azimuthal slab spectrum, the largest scale $\kappa=1$ corresponds to smallest wavenumber $k_0=1/r$ where $r$ is the heliocentric distance of the point where the spectrum is evaluated, and for the radial slab spectrum the smallest wavenumber is $k_0=\sqrt{r/a}$. 

Turbulence parameters, required for our model, can vary substantially from solar wind stream to another, and also radially, as discussed in more detail in Section~\ref{sec:discussion}. In the current study, the first one to use the new turbulence model introduced in this paper, we select one set of turbulence parameters, as given below, and leave a more in-depth investigation of the parameter space to a future publication.

The turbulence total energy varies as a function of heliocentric distance. Here we assume a power-law dependence $\delta B^2\propto r^{-\gamma}$. The WKB description of the turbulence evolution \citep{Richter1974, TuPuWei1984} suggests $\gamma=3$ in the interplanetary space, and recent modelling by \citet{Adhikari2020} based on the PSP observations gives similar values, with $\gamma=3.14$ and $2.19$ for the 2D and slab components, respectively. On the other hand, Helios observations \citep{Bavassano1982JGR} and modelling \citep{Chhiber2019PSPpreds} suggest that the ratio $\delta B/B$ remains almost constant within 1 au, indicating $\gamma=4$ close to the Sun. We choose an intermediate value of $\gamma=3.3$ for this study. Our turbulence amplitude is set by value $\delta B^2/B^2=0.03$ at 1 $\rsun$, which results in $\delta B^2/B^2\approx 0.6$  at 1~au, a value in line with observations \citep[e.g.][]{Bavassano1982JGR, Bieber1994}. The energy balance between the 2D and slab modes is set to give 80\% of the turbulence energy to the 2D modes, following \citet{Bieber1996}.

The spectral shape is formed by the Kolmogorov inertial scale, shown in Eqs.~(\ref{eq:2dspectrum}),~(\ref{eq:radslabspectrum}) and~(\ref{eq:azslabspectrum}), with a so-called energy-containing range with spectrum $k^p$ at low wave numbers. The spectral index $p$ has been reported to have an effect on particle diffusion coefficients as calculated for different particle transport theories \citep[e.g.][]{Shalchi2010_apss, Chhiber2017CRdiffGlobmodel}. We will investigate the the influence of $p$ on the field line random walk in Parker spiral in a future study. In this study, we use the power law index $p=0$ between the minimum wave number $\kappa_0=1$ and the breakpoint scales $\lambda_{c\perp}$ and $\lambda_{c\parallel}$ for the 2D and slab spectra, respectively.  

The spectral break scale lengths $\lambda_{c\parallel},$ and $\lambda_{c\perp}$ are also typically considered to have a radial dependence. Based on modelling in \citet{Chhiber2017CRdiffGlobmodel}, we will use $\lambda_{c\perp}=0.04 (r/r_\odot)^{0.8} r_\odot$, which is a reasonable fit to their correlation scale near equator, taking to account the relation between correlation length and spectral break scale $\lambda_{corr\perp}=0.8\lambda_{c\perp}$ for our spectral shape \citep{Matthaeus2007, Chhiber2021_randomwalk}. The parallel spectral breakpoint is taken as 2 times $\lambda_{c\perp}$, following similar ratios reported for the parallel and perpendicular correlation scales \citep[e.g.][]{OsmanHorbury2007}.

Finally, we set the solar wind speed to 400 km/s, the solar rotation rate is $2.86533\times 10^{-6} \;\mathrm{rad}\; \mathrm{s}^{-1}$, and the magnetic field at 1~$\rsun$ is set to 1.78 gauss.

\section{Results}\label{sec:results}

\begin{figure}
  \plotone{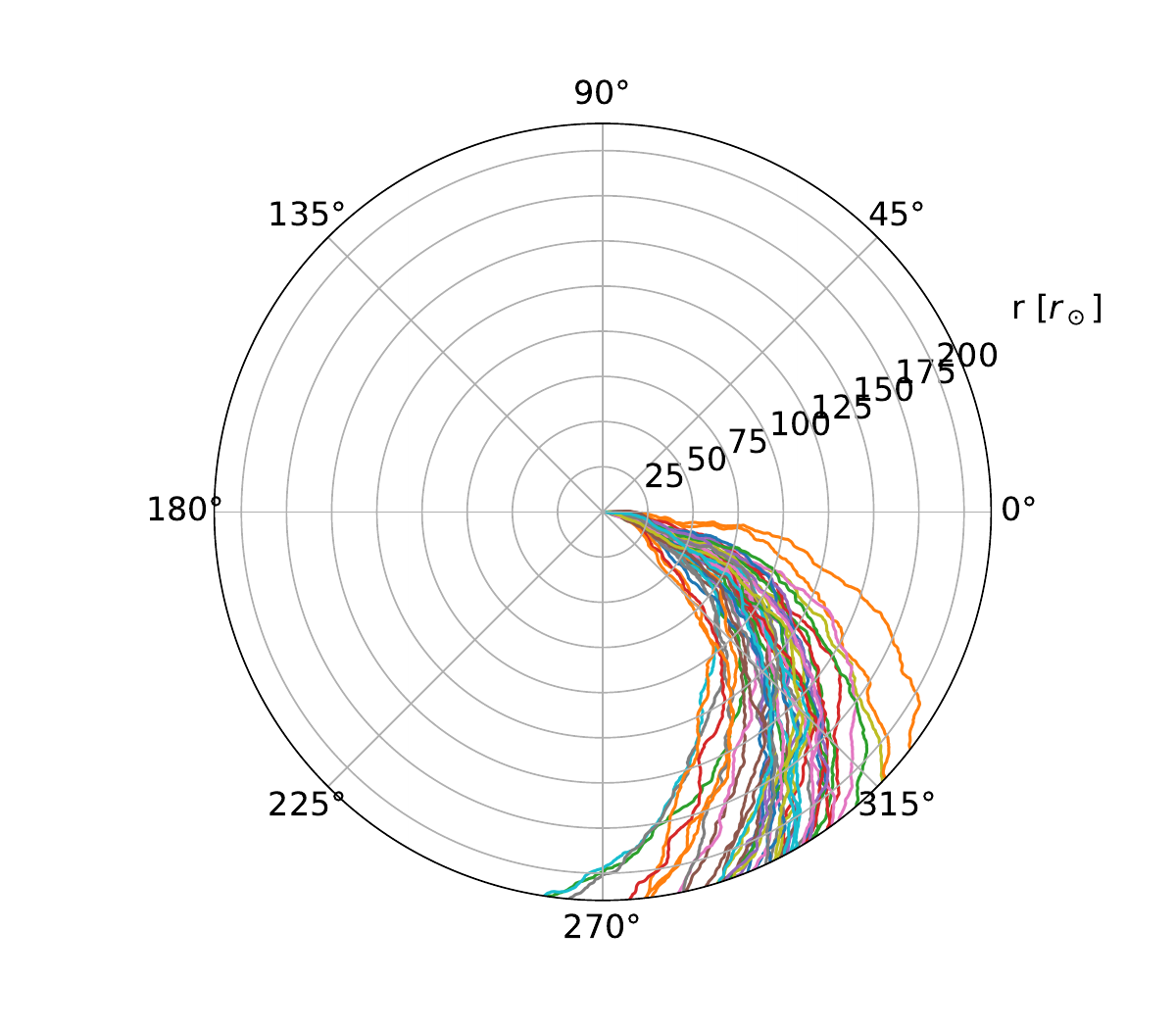} 
  \caption{A sample of field lines starting from an $8^\circ\times 8^\circ$ source region at the solar equator at 2 solar radii, for composite turbulence.}\label{fig:samplelines}
\end{figure}

\subsection{Field line maps at 1~au}

\begin{figure*}
 \includegraphics[width=0.46\textwidth]{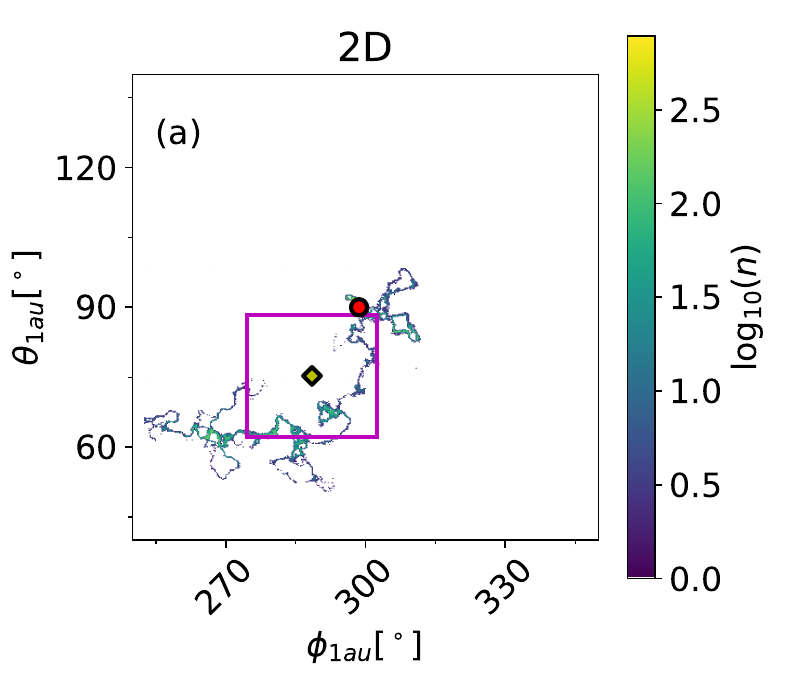}
 \includegraphics[width=0.46\textwidth]{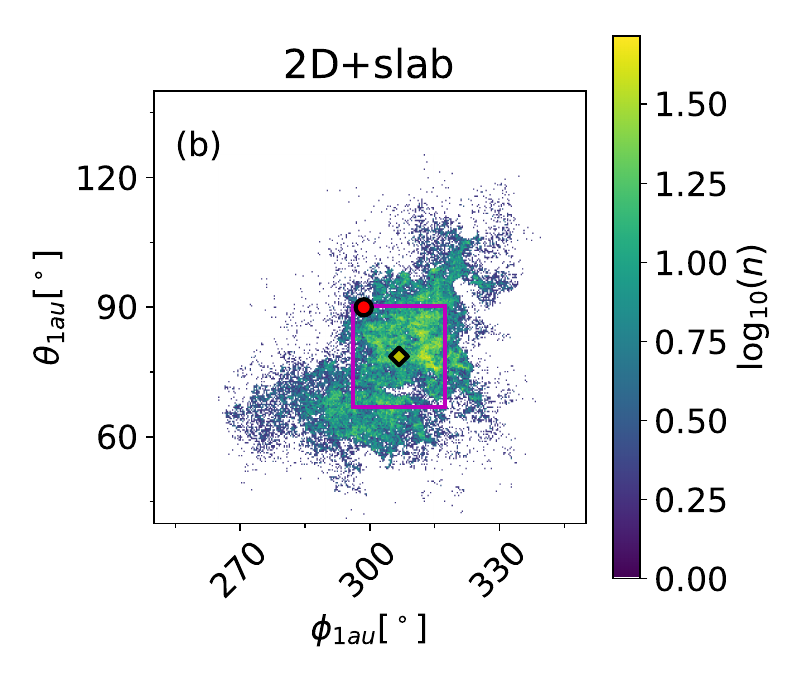}

\caption{Distribution of field line crossings at 1 au for (a) 2D turbulence; (b) composite source, for an $2\times 2^\circ$ source centred at solar equator at longitude $\phi=0^\circ$. The red circle depicts the location where the Parker spiral from the centre of the source region traverses the 1 au square. The magenta box depicts the range of standard deviation in latitude and longitude around the mean (the yellow diamond).
\label{fig:phitheta_2dcomp}}
\end{figure*}

In this section, we investigate the magnetic field line behaviour in the interplanetary space in the presence of turbulence as described by our model. We solve Equation~(\ref{eq:fieldline}) for a sample of 100,000 field lines for each simulation run. In Figure~\ref{fig:samplelines} we trace 50 field lines that start at $r=2 r_\odot$ in a latitudinal and longitudinal area of $8^\circ\times8^\circ$ centred at the solar equator at heliolongitude $\phi=0^\circ$. As can be seen, the field lines in general follow the Parker spiral geometry, however they clearly show random fluctuations, which spread the fieldlines in heliolongitude to a range considerably wider than the initial $8^\circ$.

We will now concentrate on the extent to which the field lines spread in heliolongitude and heliolatitude as they advance from the Sun to 1~au. In Figure \ref{fig:phitheta_2dcomp}, we show in panel (a) how the field lines map from $2\rsun$ a to the 1-au sphere for the 2D turbulence case displayed in Figure~\ref{fig:vecpot_phitheta}, over a wider heliolongitudinal and colatitudinal view. The colour scale shows the density of field lines in logarithmic scale. The red circle and yellow diamond show the Parker spiral crossing point and the mean of the field line crossings, respectively, and the magenta square shows the standard deviation range in latitude and longitude, centred around the mean of the field line distribution. As we can see, the field lines cross the 1-au sphere only at narrow lanes in the $\phi-\theta$ plane, spanning an angular range of 60$^\circ$ in longitude and $45^\circ$ in latitude. 

Compared to Figure~\ref{fig:phitheta_2dcomp}~(a), in Figure~\ref{fig:phitheta_2dcomp}~(b), the turbulence parameters have been changed from 100\% 2D contribution to 80\%:20\% 2D-slab mixture. As can be seen, the slab component makes it possible for the fieldlines to spread from the vector potential equipotential surfaces. As a result, the field lines map over a wider range, filling a large proportion of a $60^\circ\times 60^\circ$ angular area, with field line density variation of an order of magnitude. We do, however, see notable fine structure and regions where the field lines have no access. Also, the mean of the crossing locations is still considerably displaced from the Parker spiral crossing point (red circle).

\begin{figure*}
\includegraphics[width=0.33\textwidth]{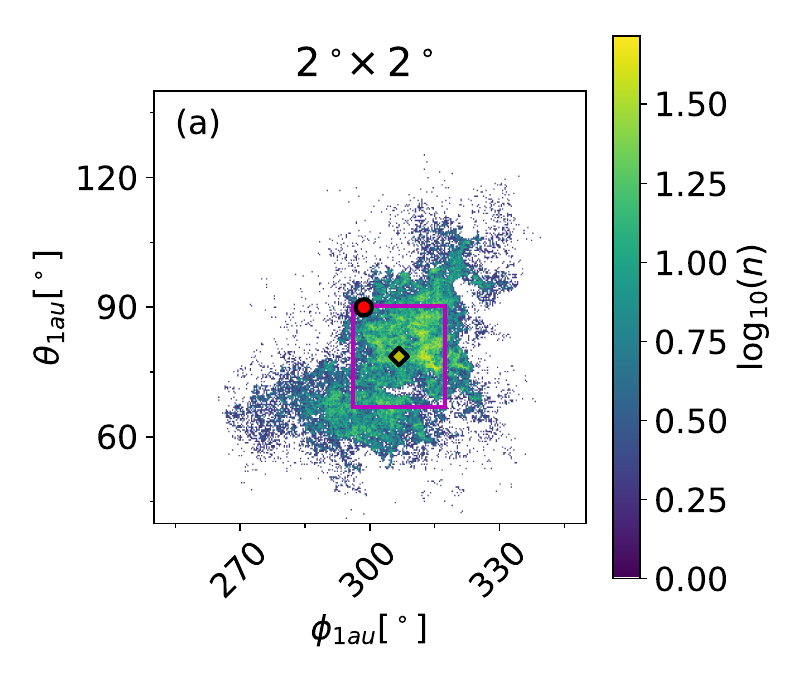} 
\includegraphics[width=0.33\textwidth]{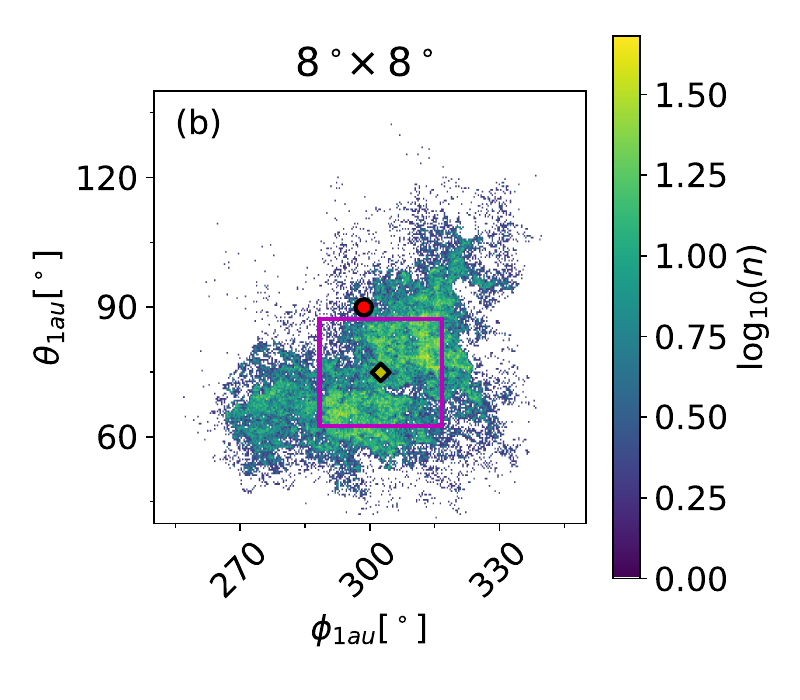} 
\includegraphics[width=0.33\textwidth]{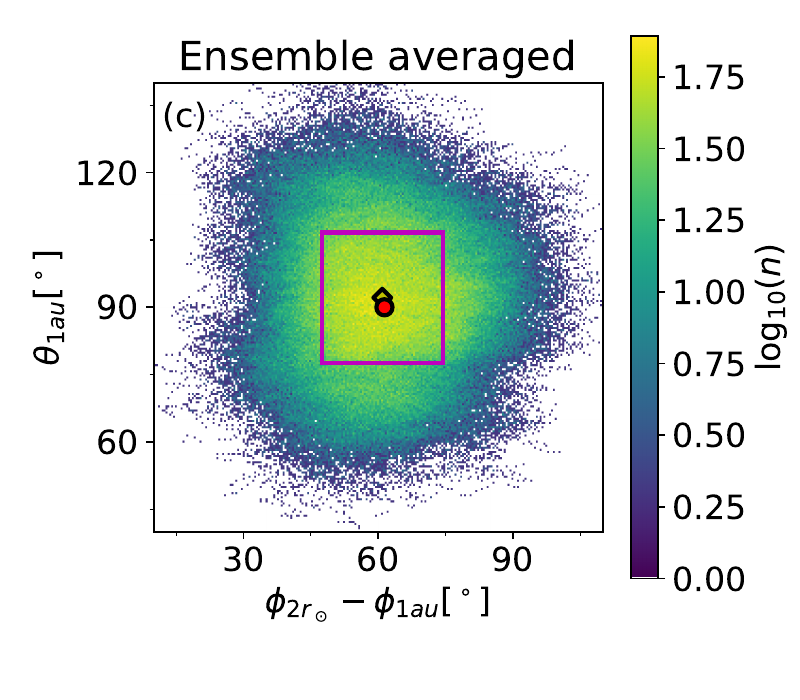} 
\caption{Distribution of field line crossings at 1 au for (a) narrow, $2\times 2^\circ$ source; (b) $8^\circ\times 8^\circ$ source, both centred at at $r=2\rsun$, $\theta=\pi/2$ and $\phi=0$, and (c) a source ensemble-averaged over longitude, with latitudinal extent of $8^\circ$, at $r=2\rsun$, $\theta=\pi/2$. In (c), the abscissa shows the difference between the initial (2~$\rsun$) and final (1~au) longitude for each field line. The symbols and the magenta box are as in Figure~\ref{fig:phitheta_2dcomp}.
\label{fig:phitheta}}
\end{figure*}

In Figure \ref{fig:phitheta} we investigate the effect of the source size on the crossing point distribution. We can see that when increasing the source size from 2$^\circ$ to 8$^\circ$, the crossing points have spread further, however, we can still see some fine structure, and notable distortions due to the local structure of the turbulent field. The longitudinal standard deviations differ significantly, with $\sigma_\phi=10.6^\circ$ for the $2\times 2^\circ$ source, and $14.2^\circ$ for the $8\times 8^\circ$ source. The latitudinal extent is of the same order as the longitudinal extent. As can be seen, the distribution of the field lines is significantly displaced from the Parker spiral (red circle). This indicates that when the source region at the Sun is small, the local turbulent structures can significantly distort the connection from the solar source to 1 au from the nominal Parker spiral.

While the local structures give information on the complexity of the field line mappings at 1~au, they complicate the comparison of the field line simulations with models based on field line diffusion, where such local structures turbulent do not exist. For this purpose, we have formed a field line crossing map averaged over an ensemble of turbulent field lines. This is done by using a wide range of source longitudes, between $0^\circ$ and $360^\circ$, and forming a density map of field line crossings at 1 au from the 1-au latitude and the change in heliolongitude of each field line from the source to 1 au, $\phi_{2\rsun}-\phi_{1\,au}$, which we present in Figure~\ref{fig:phitheta}~(c). As can be seen, the distribution is now almost symmetric with respect to the Parker spiral crossing point (the red circle). The longitudinal and latitudinal standard deviations are both about $14^\circ$, and the distribution at 1~au is centred around the Parker spiral crossing point. This kind of ensemble-averaged distribution can be more easily used for comparison of our results against field line transport models.

\subsection{Field line random walk from 2 $\rsun$ to 1~au}

In this section, we investigate how the extent of the random-walking field lines perpendicular to the mean magnetic field evolves as a function of heliocentric distance, and compare our simulation results with theoretical approaches on the field line behaviour due to plasma turbulence. The random walk of field lines is often described in terms of field line diffusion. \citet{Matthaeus1995} introduced a link between the turbulence properties and a field line diffusion coefficient for the field lines across the mean field direction in two-component turbulence as 
\begin{equation}
    D_{\perp}=\frac{1}{2}\left(D_{slab}+\sqrt{D_{slab}^2+4 D_{2D}^2}\right)
\end{equation}
where $D_{slab}$ and $D_{2D}$ are contributions to the diffusion coefficient from the slab and 2D turbulence. The contribution of the 2D turbulence can be written in the form
\begin{equation}\label{eq:diff_DD}
    D_{2D}=\tilde\lambda \frac{\sqrt{dB_{2D}^2/2}}{B}, 
\end{equation}
where $\tilde\lambda^2=\int S_\perp(k_\perp)k^{-2}d^2k/\delta B_{2D}$ is the so-called ultrascale \citep[see also][]{Matthaeus2007}, with $S_\perp(k_\perp)$ is the 2D turbulence spectrum and $k_\perp$ the perpendicular wave number. We calculated the ultrascale for our spectrum, using a piecewise approximation as in \citet{Matthaeus2007}, to find $\tilde\lambda\approx 1.34 \lambda_{c\perp}$ at the limit of small $k_0 \lambda_{c\perp}$. It should be noted that on the limit of $k_0=0$ the ultrascale diverges for $p=0$ \citep[see also][]{Engelbrecth2019}. The slab contribution, $D_{slab}=\lambda_{corr,\parallel} dB_{slab}^2/B^2$ is small compared to the 2D contribution for our parameters.

An alternative formulation for the field line diffusion coefficient was suggested by \citet{Ghilea2011}, who employed random ballistic correlation, RBD, instead of dynamic decorrelation, DD used in \citet{Matthaeus1995}, to obtain $D_\perp=D_{slab}+D_{2D,RBD}$, where
\begin{equation}\label{eq:diff_RBD}
    D_{2D,RBD}=\lambda_{corr\perp} \frac{\sqrt\pi}{2}\frac{\sqrt{dB_{2D}^2}}{B}
\end{equation}
with $\lambda_{corr\perp}=0.8\lambda_{c\perp}$ for our turbulence spectrum \citep{Chhiber2021_randomwalk}.

\begin{figure}
  \plotone{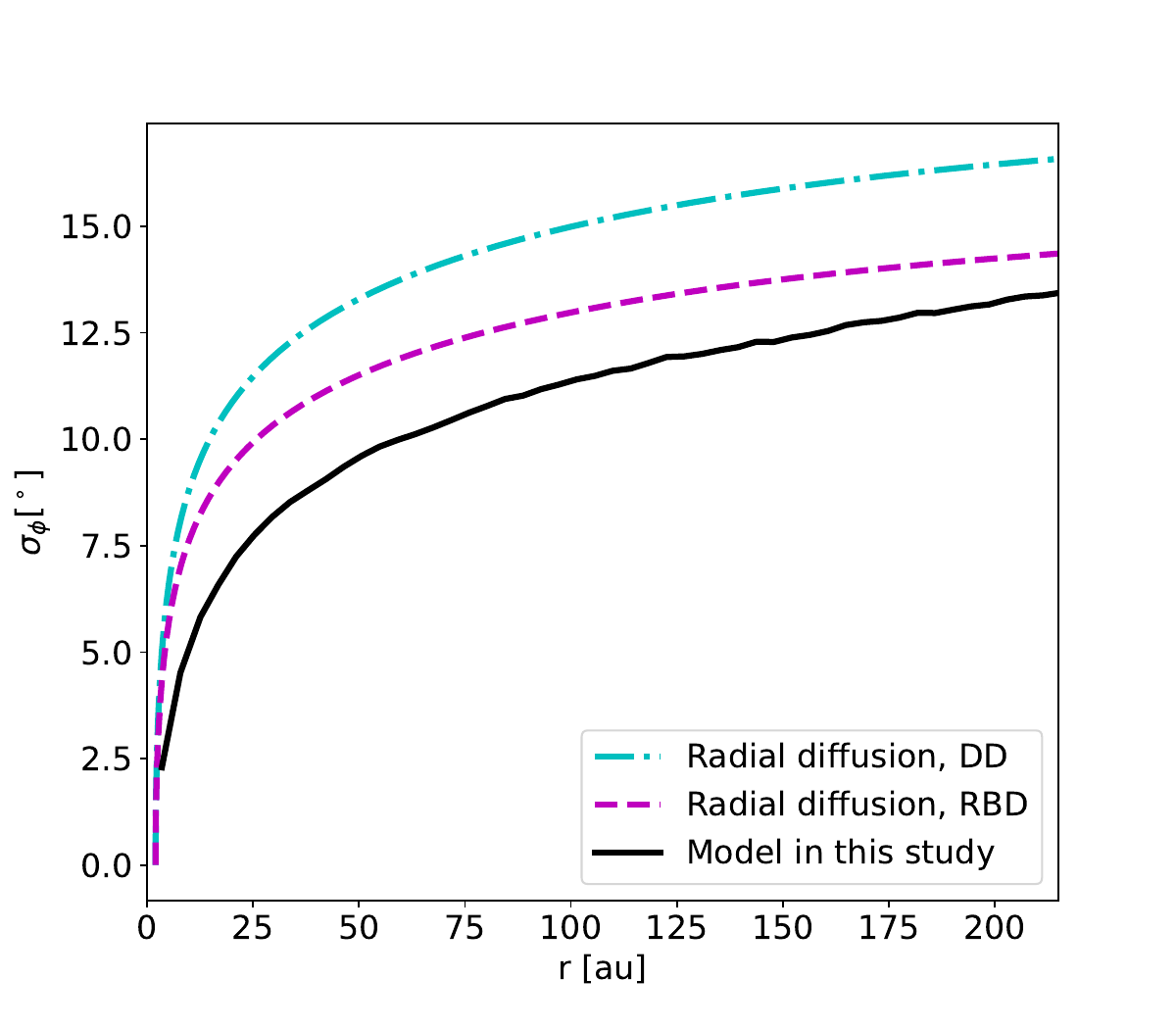} 
  \caption{The standard deviation of the longitudinal distribution of field lines as a function of distance from the Sun. The solid black curve shows the result of field line integration with our new turbulence model, whereas the dash-dotted cyan and dashed magenta curves show the solution of Equation~(\ref{eq:ode_variance})) for the DD (Equation~(\ref{eq:diff_DD})) and RBD (Equation~(\ref{eq:diff_RBD}) diffusion coefficients, respectively.\label{fig:std_comparison}}
\end{figure}

These diffusion coefficients can be used to evaluate the spread of field lines in the heliosphere. In this work, we do not solve the full field line diffusion equation in the Parker spiral. Instead, we derive in Appendix~\ref{sec:longvar} a method to directly calculate the longitudinal variance of diffusively spreading field lines in radial geometry.

In Figure~\ref{fig:std_comparison}, we present a comparison of the longitudinal standard deviation $\sigma_\phi$ of the field lines for the integrated field lines from our turbulence model (the solid black curve), and the solution of the radial diffusion Equation~(\ref{eq:ode_variance}) for the two diffusion coefficients given by Equations~(\ref{eq:diff_DD}) and~(\ref{eq:diff_RBD}) (the dash-dotted cyan and dashed magenta curves, respectively). As can be seen in the figure, the longitudinal width obtained with our new model increases rapidly within the first 30 solar radii, and then continues to increase at a much slower rate, reaching  $13.5^\circ$ at 1 au (215~$r_\odot$). Similar initial fast and later slower increase is also seen for both of the radial diffusion models. The RBD diffusion coefficient given by Equation~(\ref{eq:diff_RBD}) is closer to the $\sigma_\phi$ obtained from our new model than the DD coefficient, with both resulting in a slightly larger longitudinal extent. This can be expected, based on the comparison of the diffusion models with turbulence simulations by \citet{Ghilea2011} (with constant background magnetic field), who show that for low slab fraction and turbulence amplitude $\delta B^2/B^2$ both RBD and DD models overestimate the spreading of the field lines.

\begin{figure}
  \plotone{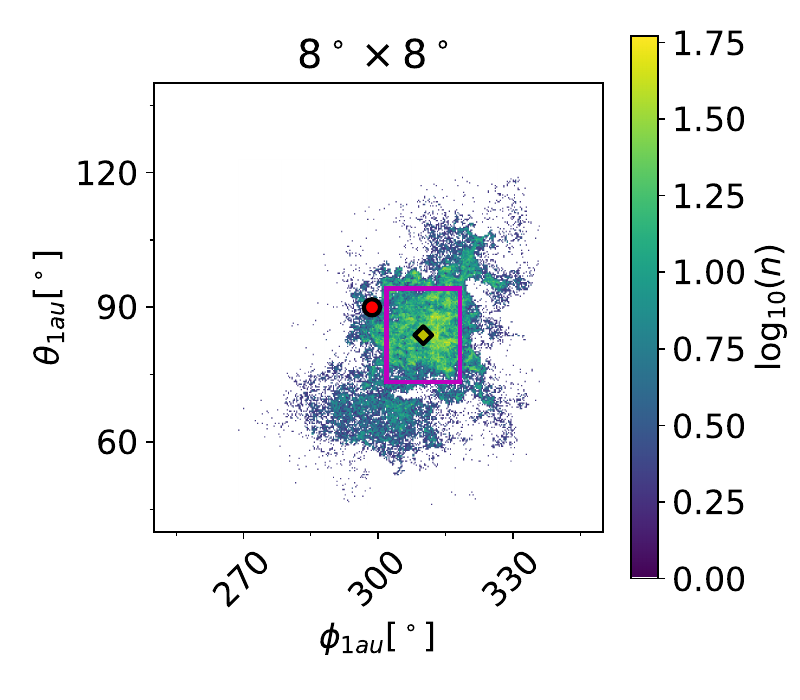} 
  \caption{Distribution of field line crossings at 1 au for an $8\times 8^\circ$ source at 20 $r_\odot$.}\label{fig:phitheta_r20}
\end{figure}

It is important to note the initial rapid increase of the longitudinal standard deviation in Figure~\ref{fig:std_comparison}, as it demonstrates that the conditions close to the Sun are very significant for the evolution of the field line extent. For the parameters in our study, the field line diffusion coefficient, as given by Equations~(\ref{eq:diff_DD}) or~(\ref{eq:diff_RBD}) scales as $D_{FL}\propto r^{1.15}$. However, it is more informative to describe diffusion in angular units, rather than distance units, as in Equation~(\ref{eq:ode_variance}), since we are interested in the evolution of longitudinal extent of the field lines. In angular units, the field-line diffusion coefficient is given as $D_{ang}\propto r^{-0.85}$, that is, a decreasing function of the heliocentric distance. Thus, the field-line random walk in longitude and latitude is dominated by the conditions at heliocentric distances close to the Sun. We demonstrate this in Figure~\ref{fig:phitheta_r20} where the field lines are integrated from an $8\times 8^\circ$ source region at $20\;\rsun$ to 1~au instead of from $2\;\rsun$. The distribution of the field line crossings at 1~au is clearly narrower than that in Figure~\ref{fig:phitheta}~(b), with longitudinal standard deviation of~$8^\circ$.

\subsection{Field line lengths in Parker spiral turbulence}

\begin{figure*}
\includegraphics[width=0.33\textwidth]{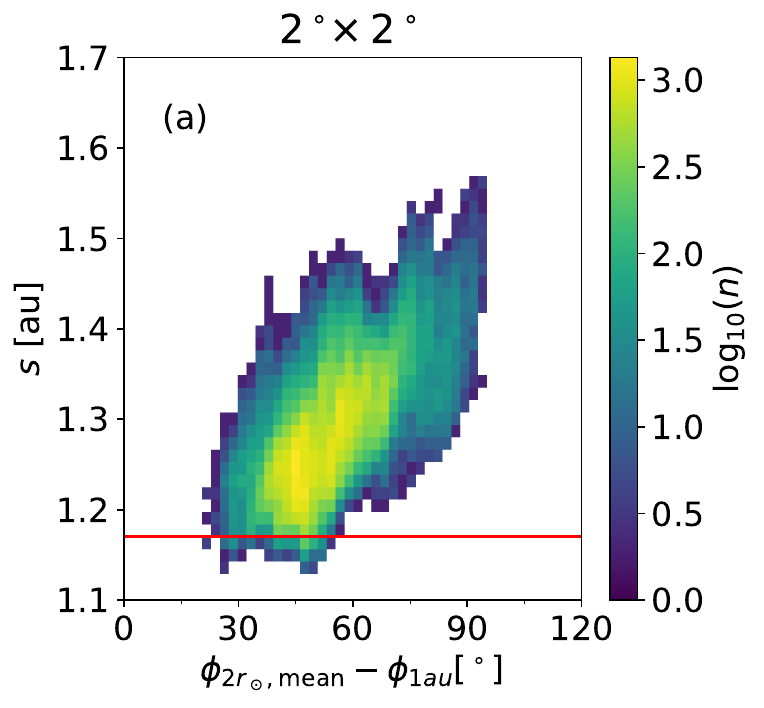} 
\includegraphics[width=0.33\textwidth]{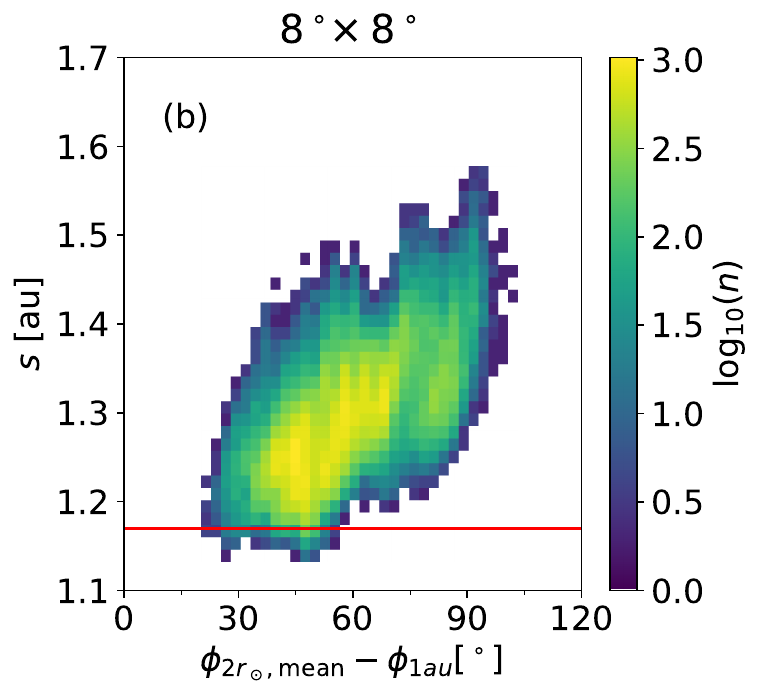} 
\includegraphics[width=0.33\textwidth]{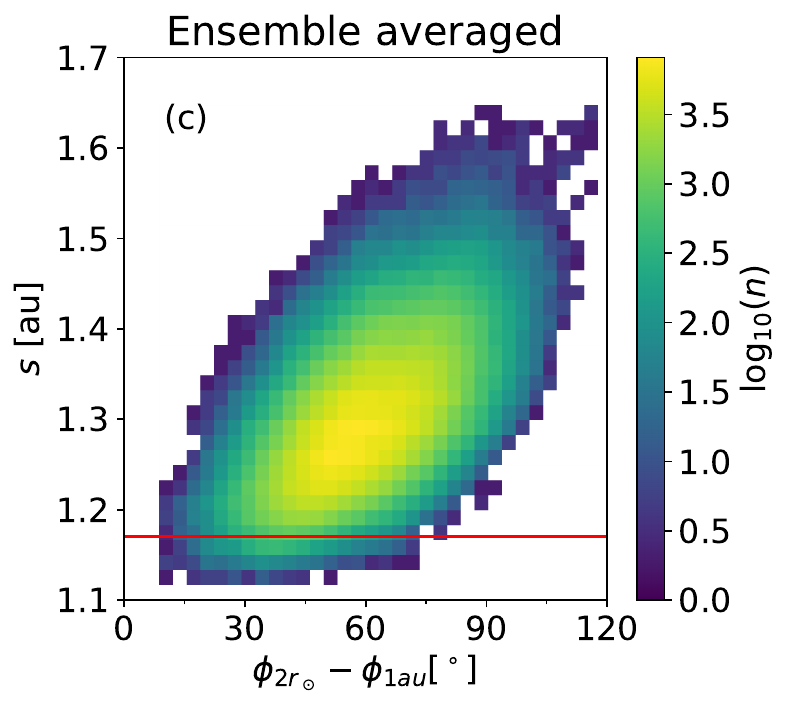} 
\caption{Contour plot of the density of the lengths of the field lines as a function of the change in field line heliolongitude from 2~$\rsun$ to 1 au for a (a) $2\times 2^\circ$ source; (b) $8\times 8^\circ$ source, and (c) source ensemble-averaged over longitudes. In (a) and (b), the abscissa shows the heliolongitude of the field line at 1~au with respect to the heliolongitude of the centre of the source region at 2~$\rsun$, (c) the distribution is formed for the change in heliolongitude of each field line as in Figure \ref{fig:phitheta}~(c). The red horizontal line at $s=1.17$~au shows the nominal Parker spiral length.
\label{fig:phipath}}
\end{figure*}

We next turn our attention to the length of the field lines and their dependence on the relative position between the source region and an 1-au observer. In Figure~\ref{fig:phipath}, we show the density of the field line lengths $s$ as a function of the change in the heliolongitude of the field line from 2~$\rsun$ to 1~au, summed over latitude. The abscissa in panels (a) and (b) is the change between heliolongitude of the centre of the source region and the field line at 1~au. In Figure~\ref{fig:phipath}~(c) the distribution is formed from an ensemble average of source longitudes as in Figure~\ref{fig:phitheta}~(c), and the abscissa is the change in the heliolongitude for each field line in the distribution.

As we can see in all of the panels of Figure~\ref{fig:phipath}, the path length in our turbulence model is typically significantly longer than the Parker spiral length, shown by the horizontal red line in the panels. At 60$^\circ$, the field line length varies between 1.15 and 1.45~au (within 1~order of magnitude of the most likely length). The field line length distribution is strongly dependent on the source longitude of the field lines. Locations with small angular separation $\phi_{2\rsun}-\phi_{1au}$, corresponding to sources that would be near the centre of the solar disk from the perspective of 1-au observer,  tend to have smaller $s$-values, close to the nominal Parker spiral length. Some field lines, east from the nominal Parker Spiral-connected longitude $\phi_{2\rsun}-\phi_{1au}\sim 60^\circ$, are shorter than the nominal Parker spiral length, due to turbulence "straightening" the curved Parker spiral to more radial field line \citep[see also][]{LaDa2019pathlength}. However, even from those source longitudes, the majority of the field lines are longer than the nominal Parker spiral.  Further to the west, the source regions are connected to the observer at 1~au by progressively longer paths, with the behind-the-west-limb sources almost 40\% longer than the nominal Parker spiral.

The effect of the source size can be seen in the three panels of Figure~\ref{fig:phitheta}. Panel~(a) refers to a small source, and we can see that the range of longitudes is patchier and narrower than in the wider source in panel~(b), with intermittent ranges of shorter and longer path lengths. In panel~(c) the field line lengths are ensemble-averaged over source longitudes, which results in smoothing over the fine structure of the small localised source regions.

\section{Discussion}\label{sec:discussion}

In this paper, we presented the first model of interplanetary turbulence in a Parker spiral geometry that contains a 2D component with respect to the Parker spiral mean field direction. The wave modes are spaced logarithmically, making it possible to model turbulent scales over several orders of magnitude. The turbulence is defined using an analytic formulation, which enables us to avoid the divergence of the field that may arise from interpolation of the field on finite grids. 
We validated the model by showing that the 2D turbulence model fulfills the required invariance of the vector potential along the Parker spiral mean magnetic field, and the trapping of field lines on constant vector potential equisurfaces (Figures~\ref{fig:vecpot_rphi} and~\ref{fig:vecpot_phitheta}). Further, we demonstrated that the evolution of the longitudinal extent of the field lines is consistent with the commonly-used theoretical descriptions, taking into account the applicability of those models within the heliospheric turbulence parameter range (Figure~\ref{fig:std_comparison}). The new model enables us to investigate the turbulent field line behaviour in an environment where the turbulence varies radially and is coupled to the non-radial, Parker spiral large-scale geometry. Applied to charged particle transport in magnetised turbulence, the model can be used to investigate the interplay between large-scale drifts, particle propagation initially along meandering field lines and, at longer timescales, diffusively across the mean magnetic field.

Our results agree with suggestions in earlier field line simulation studies that find the field lines spread to a large range of longitudes and latitudes in the heliosphere \citep[e.g.][]{Tooprakai2016,Chhiber2021_randomwalk}\footnote{Note that \citet{GiaJok2004} footpoint random walk model results in narrower range}. We also note that the regions with depleted field line density in the field line maps (Figure~\ref{fig:phitheta}~(a) and~(b)) may offer an explanation for energetic particle intensity dropouts reported in some SEP events \citep[e.g.][]{Mazur2000}, as already reported earlier simulation studies \citep{GiaJokMaz2000, Ruffolo2003,Tooprakai2016}.

The wide heliolongitudinal and -latitudinal extent of field lines, due to turbulence, is significant also for the observed wide longitudinal extent of solar energetic particle events. Several works report a wide longitudinal extent of the peak SEP intensities, with standard deviation 30-50$^\circ$ \citep{Lario2006, Lario2013, Wiedenbeck2013, Cohen2014, Richardson2014}. This range is wider than the field-line extents in our study, with longitudinal standard deviation of 13$^\circ$, however a wider range could be obtained with stronger turbulence, particularly closer to the Sun. The wider SEP peak intensity extent may also be affected by cross-field diffusion of particles that spreads the SEPs further from the meandering field lines \citep[see e.g.][]{LaEa2013b, LaEa2017meandstatistic}. It should be noted that the longitudinal extent of SEPs can be due to several mechanisms, such as wide SEP sources at CME-driven shock waves \citep[e.g.][]{Cliver1995_CME_SEP_source}, field line spreading in the corona \citep[e.g.][]{Liewer2004_coronalopening}, sympathetic flaring \citep[e.g.][]{Schrijver2011_sympathetic}, and cross-field diffusion from the mean magnetic field (as opposed from the turbulently meandering field lines \citep[e.g.][]{Zhang2009, Droge2010, Strauss2017perpel} \citep[see also, e.g.,][for discussion]{Wiedenbeck2013}. Thus, our results only strengthens the suggestions that field-line random walk is one possible mechanism, without overruling other mechanisms. We will investigate this issue by performing full-orbit simulations in the magnetic field configuration described in this paper in near-future.

\begin{figure}
  \plotone{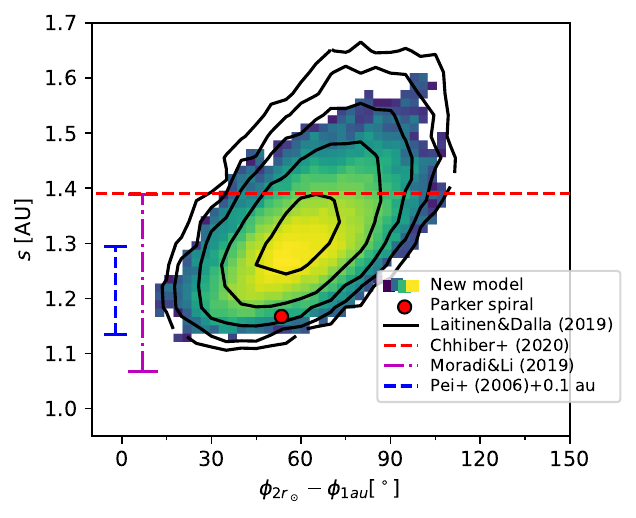} 
  \caption{The length of the field line between $2 \rsun$ and 1~au, as a function of longitude at 1~au. The colour contour distribution is formed for the change in heliolongitude as in Figure \ref{fig:phipath}~(c).\label{fig:phipath_vs_stoch}}
\end{figure}

Magnetic field line lengths to 1~au have recently been investigated by several research groups, using a variety of different modelling approaches. \citet{Chhiber2020Pathlengths} introduced a simple method based on random-walking field lines to evaluate length of turbulent field lines with  $s/s_0=B/B_0\approx\sqrt{1+\delta B^2/B^2}$ where $B_0$ and $B$ are magnitudes of the background and the total magnetic field, $\vect{B}=\vect{B}_0+\vect{\delta B}$, and $s_0$ the length of the undisturbed field line. The red horizontal dashed line in Figure~\ref{fig:phipath_vs_stoch} shows this estimate for the $\delta B^2/B^2$ used in our study (see Section~\ref{sec:turbparams}). The obtained path length of \citet{Chhiber2020Pathlengths} is slightly larger than the peak of our simulations (filled contour), possibly because their simple model does not evaluate the sum of random-walking step lengths as a stochastic process. They also propose a more rigorous method to evaluate the path length, but due to its complexity for non-constant $\delta B/B$ heliosphere, we do not present that estimate here. \citet{Chhiber2020Pathlengths} also analysed field-line simulations in composite turbulence in radial geometry, which in general agreed with both their simple and rigorous analytic models.

The black contour lines in Figure~\ref{fig:phipath_vs_stoch} show the path length distribution derived from the stochastic model by \citet{LaDa2019pathlength}, adjusted to the parameters of our study (including using the RDB field line diffusion coefficient \citep{Ghilea2011} instead of DD \citep{Matthaeus1995}, see Appendix~\ref{sec:app-stoch-field-line}). As one can see, both the range of path lengths and the heliolongitudinal dependence of the path lengths obtained with the \citet{LaDa2019pathlength} method is very similar to that in our study. It should be noted that the \citet{LaDa2019pathlength} model is 2-dimensional, which is likely to result in shorter path lengths than a 3-dimensional model would. Also, as implied by Figure~\ref{fig:std_comparison}, the field line diffusion coefficient given by Equation~(\ref{eq:diff_RBD}) is larger than that implied by our simulation studies, resulting in longer path lengths. These two effects compensate each other to some extent, which results in good agreement with the results from our new analytical turbulence model.

The vertical bars in Figure~\ref{fig:phipath_vs_stoch} show the ranges obtained by \citet{Pei2006} (dashed blue) and \citet{MoradiLi2019} (dot-dashed magenta) using the magnetic field footpoint motion model of \citet{Giacalone2001} for random-walk rms speeds 4~km/s and 2.5~km/s, respectively. The values are obtained as range where the field line length distribution $f(s)$ fulfills the condition  $f(s)>\mathrm{max}\{f(s)\}\,\mathrm{e}^{-1/2}$. The \citet{Pei2006} study started their field lines from 0.1~au (which we have compensated in Figure~\ref{fig:phipath_vs_stoch}), which may contribute to the lower range of path lengths for 4~km/s RMS speeds, as compared to the \citet{MoradiLi2019} 2.5~km/s case. It should be noted that the \citet{Giacalone2001} model turbulence is not purely transverse with respect to the Parker spiral, and this may affect the path lengths, as compared to our path lengths. Direct comparison of the \citet{Pei2006} and \citet{MoradiLi2019} path lengths with our model is not trivial, as the spectral shape and the range of spectral scales in their works are very different from that used in our study. 

The length of the field lines from the Sun to Earth is often discussed in the context of the arrival of SEPs from their sources near the Sun. A popular method for obtaining the solar particle injection time, the velocity dispersion analysis (VDA) method \citep[e.g.][]{Lin1981} fits the observed SEP onset times and returns the solar injection time and the path length the first particles have travelled, under assumption that the first particles have propagated scatter-free. The obtained path lengths are often very long compared with the nominal Parker spiral lengths (e.g., in large-statistic studies of SOHO, ACE and STEREO-observed SEP events \citep{Paassilta2017,Paassilta2018} and analysis of Ulysses SEP observations \citep{DallaEa2003Annales}). Recently, long VDA path lengths have been reported also close to the Sun \citep[e.g.][]{LeskeEa2020}. However, the VDA method is known to produce unreliable results particularly for the path length due to its assumption of scatter-free propagation of first particles, which essentially ignores the effect of SEP transport in turbulent medium. Several modelling studies have shown that the transport effects have a significant effect on the VDA path lengths \citep[e.g.][]{Lint04,Saiz05,Laitinen2015onset,Wang2015}. However, our results \citep[and that of, e.g.][]{Pei2006,LaDa2019pathlength,MoradiLi2019,Chhiber2020Pathlengths}, suggest that the lengthening may be caused also by physical effects instead of only being a byproduct of ambitious assumptions within the VDA method.

As discussed in \citet{LaDa2019pathlength}, not properly accounting for the field line lengths may have significant implications on past modelling efforts of SEP transport that include the transport as spatial diffusion across the mean field. Such models are unphysical in that the spreading of the particle population diffusively across the mean field is not constrained by causality \citep[see also][]{Strauss2015}. This can be readily seen in the Stochastic Differential Equation (SDE) description of the diffusion equation \citep[e.g.][]{Gardiner2009}: only the propagation along the mean field direction is consistent with the particle velocity, whereas the stochastic displacement across the mean magnetic field contributes to the distance propagated by the pseudo-particle, but takes no time. This can result in non-physical arrival times for the modelled SEPs. In particular the asymmetry of the path lengths caused by the Parker spiral geometry (as seen in Figure~\ref{fig:phipath}) may significantly affect the SEP time-intensity profiles.

The asymmetry caused by the Parker spiral has also other consequences for energetic particles. Charged particles experience latitudinal drifts due to the gradient and curvature of the Parker spiral field, resulting in deceleration particles in the convective solar wind electric field. \citep[e.g.][]{Marsh2013, Dalla2013, Dalla2015}. However, it is unclear how these drifts interplay with the meandering of the field lines. Comparison of galactic cosmic ray observations with modulation models \citep[e.g.][]{PotgieterEa1989} suggests that the heliospheric charged particle drifts, as described by the antisymmetric part of the cosmic ray diffusion tensor \citep{JokipiiEa1977} are reduced, and it has recently been suggested that similar reduction would affect also the drifts of solar energetic particles \citep[e.g.][]{Engelbrecht_2017,Vandenberg_2021}. Simulation studies have indicated that turbulence would have the effect of reducing the drifts \citep[e.g.][]{Giacalone_1999_drifts, Minnie2007_drifts, Tautz2012_drifts}. It should be noted that \citet{Giacalone_1999_drifts} and \citet{Tautz2012_drifts} investigate the non-zero asymmetric part of the diffusion tensor in the presence of constant background magnetic field, that is, in a configuration where macroscopic drifts do not exist \citep[see also][]{BurgerVisser2010}, whereas \citet{Minnie2007_drifts} implements a non-constant background that produces macroscopic gradient drift, however without curvature or spatially varying turbulence parameters. Our new turbulence model provides an ideal tool for investigating the drift reduction of solar energetic particles in the heliospheric environment with curvature and gradient drifts in a spatially varied turbulence, and this will be the subject of a future study.

The path lengths and the longitudinal and latitudinal extent of the field mapping at 1~au depend on the turbulence parameters and their spatial dependence in the heliosphere. In this study, we used moderate turbulence parameters, with the energy-containing range spectral index $p=0$, $\lambda_{corr\perp}\propto r^{0.8}$ and $\delta B^2\propto r^{-3.3}$. The values used for these parameters vary greatly in the literature. The energy-containing range spectrum affects the relation of the correlation scale and the spectral break scale, and through that connection also the field-line and particle diffusion coefficients \citep[e.g.][]{Matthaeus2007, Shalchi2010_apss, Engelbrecht2015}. For correlation length, one option is to use the \citet{Hollweg1986} assumption of $\lambda_{corr}\propto 1/B^{1/2}$ giving $\lambda_{corr}\propto 1/r$ close to the Sun \citep[e.g.][]{Perri2020}, whereas some studies have used constant correlation lengths \citep[e.g.][]{QinWang2015,LaEa2016parkermeand,Strauss2017perpel}, and some use results of turbulence transport simulations \citep[e.g][]{Chhiber2017CRdiffGlobmodel,Vandenberg_2021}. Recent results from Parker Solar Probe \citep[PSP,][]{Fox2016_PSPpaper} suggest radial dependence of $\lambda_{corr}\propto r^{0.746}$  at heliocentric heights above 30~$\rsun$, close to our model assumption. It should be also noted that the PSP observations suggest $\lambda_{corr}$ to vary an order of magnitude from solar wind stream to another \citep{Chhiber2021-PSPorbits}. Further, the ratio of the slab and 2D correlation lengths can vary both from solar wind stream to another, and radially \citep[e.g.][]{Weygand2011, Adhikari2022_turbaniso}.

The amplitude of the fluctuations, $\delta B^2$, and its dependence on distance from the Sun, also vary considerably, both in observations and in modelling. The Helios observations found that  $\delta B^2/B^2$ remained almost constant between 0.3 and 1~au in some frequency ranges \citep{Bavassano1982JGR}, varying up to an order of magnitude at different times and solar wind streams \citep[see also][for recent PSP observations]{Chhiber2021-PSPorbits}, and this near-constancy has been used in several field line and cosmic ray modelling studies, with values up to $\delta B/B\sim 1$ \citep[e.g.][]{QinWang2015,Chhiber2020Pathlengths}. However, as discussed in Section~\ref{sec:turbparams}, there is also significant support for WKB turbulence evolution, with $\delta B^2\propto 1/r^3$ in the interplanetary space, which would result in a radially-increasing $\delta B^2/B^2$ closer to the Sun. It should be noted that close to the Sun, in sub-Alfv\'enic regime, the WKB approach gives a weaker radial dependency, depending on the solar wind model \citep[e.g. $\delta B^2\propto 1/r$ in sub-Alfv\'enic WKB solar wind in][]{LaEa2016parkermeand}. Finally, also the energy partition between the slab and 2D modes, here 20:80\%, has been reported to vary in different solar wind environments \citep[see][for review]{Oughton2015} and also radially \citep{Adhikari2022_turbaniso}.

In a future study, we will investigate how the different models and observed ranges of the turbulence parameters, particularly close to the Sun in the sub-Alfv\'enic solar wind, affect the field-line behaviour in turbulent interplanetary space. While our understanding on the sub-Alfv\'enic region is limited, the recent and future observations performed by PSP \citep[e.g.][]{Zhao2022_subsuperalfturb, Bandyopadhyay2022_PSPsubalf} will guide us to better understanding of turbulent heliospheric field lines.

\section{Conclusions}\label{sec:conclusions}

In this paper, we have investigated the turbulent field line behaviour in the heliosphere, where the mean magnetic field is in a Parker spiral shape. The main contributions from our work are:
\begin{enumerate}
    \item  We have introduced a new analytical model for interplanetary turbulence, where the dominant 2D waves are transverse with respect to the Parker spiral magnetic field direction, and the minor slab component is approximated with a inner-heliosphere radial and outer-heliosphere azimuthal slab component.
    \item For the interplanetary turbulence parameters in our study, the magnetic field traced from an $8^\circ\times 8^\circ$ heliolongitudinal range at the equator at 2~$\rsun$ maps to a heliolatitudinal and -longitudinal range of approximately $60^\circ\times 60^\circ$ area at 1~au heliocentric distance, with standard deviation $14^\circ$. Areas with smaller field line density are found, in particular for smaller source regions, indicating potential regions where SEP droupouts could take place. 
    \item The radial evolution of the longitudinal and latitudinal extent of the mapped field lines is slower than predicted by two popular field line diffusion models  \citep{Matthaeus1995,Ghilea2011}, consistent with \citet{Ghilea2011} comparison of the models with simulations in cartesian geometry for the turbulence parameters used.
    \item The turbulence parameters at small heliocentric distances are central to the longitudinal and latitudinal extent of the mapped field lines, and understanding the turbulence composition low in the corona is of utmost importance for understanding the causes behind solar energetic particle event cross-field extents.
    \item The lengths of the field lines from 2~$\rsun$ to 1~au are significantly longer than that of the Parker spiral magnetic field. Further, field lines connecting 1-au observer to behind-the-west-limb sources are longer than those connecting the observer to on-disk sources. Our model, and that presented in \citet{LaDa2019pathlength}, provide a useful tool for the interpretation of solar energetic particle onset observations. 
\end{enumerate}

Our results imply that the diffusion-based SEP transport models can result in erroneous solar injection times of the SEPs, as they are not able to account for the increased length of the random-walking field lines. In future work, we will employ full-orbit simulations of energetic particles in the newly modelled interplanetary turbulence to analyse this effect, and compare the full-orbit simulations to predictions of the diffusion models.
\\


TL and SD acknowledge support from the UK Sci-
ence and Technology Facilities Council (STFC) through
grant and ST/V000934/1.
CW and SD acknowledge support
from NERC via the SWARM project, part of the
SWIMMR programme (grant NE/V002864/1).
This work was performed using resources provided
by the Cambridge Service for Data Driven Discovery
(CSD3) operated by the University of Cambridge Re-
search Computing Service (www.csd3.cam.ac.uk), pro-
vided by Dell EMC and Intel using Tier-2 funding from
the Engineering and Physical Sciences Research Coun-
cil (capital grant EP/P020259/1), and DiRAC fund-
ing from the Science and Technology Facilities Council
(www.dirac.ac.uk).
TL acknowledges support from the International
Space Science Institute through funding of the Inter-
national Team \#35 "Using Energetic Electron And Ion Observations To Investigate Solar Wind Structures And Infer Solar Wind Magnetic Field Configurations".

%

\vspace{5mm}


\software{ 
    WebPlotDigitizer \citep{Rohatgi2022_Webplotdigitizer},
    vaex \citep{Vaex},
    SymPy \citep{sympy}
    }



\appendix

\section{Equation for longitudinal variance in radial geometry}\label{sec:longvar}

Here we derive an equation for longitudinal variance for diffusively propagating passive scalars, to aid in estimation of the spread of fieldlines for different radial dependencies of the fieldline diffusion coefficient. We start with diffusion-convection equation
\begin{equation}
    \pder{f}{t}+\nabla\cdot\left(\vect{v} f\right) = \nabla\cdot\left(D\nabla f\right) 
\end{equation}
where $f$ is the density of the passive scalars, $v$ their velocity and $D$ the diffusion coefficients. We limit to the 2D case in $r-\phi$ plane, with velocity describing the radial motion, and $D$ the longitudinal diffusion. Further, we write $D_{FL}=D/v$. Going to steady-state limit, we can thus write the equation in form
\begin{equation}\label{eq:steadyfldiff}
    \frac{1}{r^2}\pder{}{r}\left(r^2f\right) = \frac{1}{r^2}\pder{}{\phi}D_{FL}\pder{f}{\phi}
\end{equation}

We next take the zeroth and second moments of Equation~(\ref{eq:steadyfldiff}). For the zeroth moment, $f_0$, the right-hand side vanishes, and the remaining equation integrates to
\begin{equation}
    f_0(r)=\left(\frac{r_0}{r}\right)^2 f_{00}
\end{equation}
where $f_{00}$ is the density at $r_0$ integrated over $\phi$. 

The second moment of the left-hand side of Equation~(\ref{eq:steadyfldiff}) is trivial. The right-hand side reduces to $2D_{FL} f_0$ at the limit $\min\{f(r, \phi)\}<<f_0(r)$. Thus, we get
\begin{equation}
    \oder{}{r}\left(r^2 f_2\right)=2 D_{FL} f_0
\end{equation}
where $f_2=\int \phi^2 f d\phi$. We can rewrite this into an ordinary differential equation for the longitudinal variance $V_\phi=f_2/f_0$ as
\begin{equation}\label{eq:ode_variance}
    \oder{V_\phi}{r}=2 D_{ang}.
\end{equation}
where we have defined the diffusion coefficient in angular units as $D_{ang}=D_{FL}/{r^2}$.
For power-law $D_{ang}= D_{ang 0}\,\left(r/r_0\right)^\alpha$, this integrates simply to 
\begin{equation}
V_\phi=V_{\phi0}+2\frac{1}{\alpha+1} D_{ang 0} \left(\frac{r}{r_0}\right)^{\alpha+1}.    
\end{equation}

\section{Stochastic field line length with random ballistic decorrelation approach to field line diffusion}\label{sec:app-stoch-field-line}

\citet{LaDa2019pathlength} derived the length of a meandering magnetic field using random walk for step length across the mean magnetic field direction comparable to the turbulence ultrascale, $\left<\Delta x^2\right>=\tilde\lambda^2$. The corresponding step length along the field, $\Delta z$ can then be obtained by using the definition of field-line diffusion coefficient, $D_{FL}=\left<\Delta x^2\right>/(2\Delta z)$. In \citet{LaDa2019pathlength}, the diffusion coefficient was taken to be the one given by \citet{Matthaeus1995}, resulting in
\begin{equation*}
\Delta z=\frac{\tilde\lambda^2}{2D_{FL}}= \tilde\lambda\frac{B}{\sqrt{2\delta B^2}}
\end{equation*}
However, in this study we found that the \citet{Ghilea2011} random ballistic decorrelation (RBD) method yields a diffusion coefficient that agrees better with our results than that of \citet{Matthaeus1995}. Using the RBD diffusion coefficient, we get 
\begin{equation}
  \Delta z_{RBD}=\frac{\tilde\lambda^2}{2D_{FL, RBD}}= \tilde\lambda\frac{B}{\sqrt{\pi\delta B^2}}
\end{equation}

The change from the \citet{Matthaeus1995} \citet{Ghilea2011} diffusion coefficient changes also the  simple approximation in \citet{LaDa2019pathlength}, to the form
\begin{equation}
\left<s(r, \phi)\right>=s_0(r, \phi)\left(1+\eta\rho\, \delta B^2/B^2\right)   
\end{equation}
where $s_0(r,\phi)$ is the length of a Parker spiral from the source longitude to longitude $\phi$, $\rho$ is a geometric factor within range $[0,1]$, and $\eta=(\pi/2)(\lambda_{corr\perp}/\tilde\lambda)^2\approx 0.56$.


\bibliography{ms}{}
\bibliographystyle{aasjournal}



\end{document}